\documentclass[aps,prl,superscriptaddress,preprintnumbers,nofootinbib,12pt]{revtex4-2}

\usepackage{graphicx,amsmath,amssymb}
\usepackage{bigints}
\usepackage{hyperref}
\usepackage{bm}
\usepackage{ulem}
\usepackage{hyphenat}

\hypersetup{colorlinks=true, linkcolor = [rgb]{0,0.08,0.45}, citecolor = [rgb]{0,0.08,0.45}, urlcolor = [rgb]{0,0.08,0.45}}

\newcommand{\req}[1]{(\ref{#1})}
\newcommand{\lb}{\label}
\newcommand{\nn}{\nonumber}

\begin{document}

\preprint{SLAC-PUB-17765}

\title{The strong coupling in the nonperturbative and near-perturbative regimes}

%\thanks{

\author{Guy~F.~de~T\'eramond}
\email[]{guy.deteramond@ucr.ac.cr}
\affiliation{Laboratorio de F\'isica Te\'orica y Computacional, Universidad de Costa Rica, 11501 San Jos\'e, Costa Rica}

\author{Arpon~Paul}
\email[]{paul1228@umn.edu}
\affiliation{School of Physics and Astronomy, University of Minnesota, Minneapolis, Minnesota 55455, USA}

\author{Stanley~J.~Brodsky}
\email[]{sjbth@slac.stanford.edu}
\affiliation{SLAC National Accelerator Laboratory, Stanford University, Stanford, CA 94309, USA}

\author{Alexandre~Deur}
\email[]{deurpam@jlab.org}
\affiliation{Thomas Jefferson National Accelerator Facility Newport News, Virginia 23606, USA}

\author{Hans~G\"unter~Dosch}
\email{h.g.dosch@gmail.com}
\affiliation{Institut f\"ur Theoretische Physik der Universit\"at, D-69120 Heidelberg, Germany}

\author{Tianbo~Liu}
\email{liutb@sdu.edu.cn}
\affiliation{Key Laboratory of Particle Physics and Particle Irradiation (MOE), Institute of Frontier and Interdisciplinary Science, Shandong University, Qingdao, Shandong 266237, China}
\affiliation{Southern Center for Nuclear-Science Theory (SCNT), Institute of Modern Physics, Chinese Academy of Sciences, Huizhou 516000, China}

\author{Raza~Sabbir~Sufian}
\email[]{gluon2025@gmail.com}
\affiliation{RIKEN-BNL Research Center, Brookhaven National Laboratory, Upton, New York 11973, USA}
\affiliation{Physics Department, Brookhaven National Laboratory, Upton, New York 11973, USA}

\date{\today}

%\collaboration{HLFHS Collaboration}

\date{\today}

%%%%%%%%%%%%%%%%%%%%%%%%%%%%%%%%%%%%%%%%%%%%%%%%%

\begin{abstract}

\vspace{2pt}
 
\begin{center}
(HLFHS Collaboration) \\
\vspace{15pt}
\includegraphics[scale=0.2]{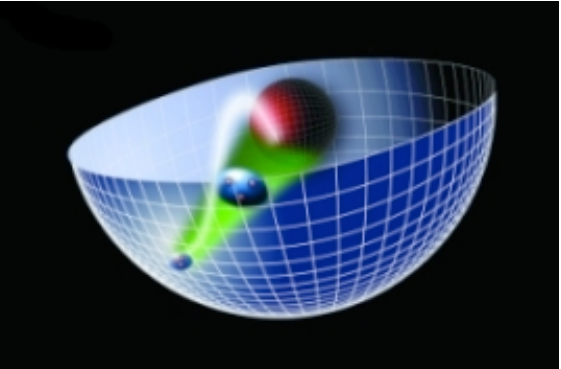}\\
\end{center}

\vspace{25pt}

We use analytic continuation to extend the gauge/gravity duality nonperturbative description of the strong force coupling into the transition, near-perturbative, regime where perturbative effects become important. By excluding the unphysical region in coupling space from the flow of singularities in the complex plane, we derive a specific relation between the scales relevant at large and short distances; this relation is uniquely fixed by requiring maximal analyticity. The unified effective coupling model gives an accurate description of the data in the nonperturbative and the near-perturbative regions.

\end{abstract}

\maketitle

\newpage

{\it Introduction.}-- The  running strong coupling $\alpha_s$  has been determined to high orders in perturbation theory at large momentum transfer: It leads to asymptotic freedom, a fundamental property of quantum chromodynamics (QCD)~\cite{Gross:1973id, Politzer:1973fx}. However, a purely perturbative description of $\alpha_s$ is inconsistent: It breaks down at the Landau pole at large distances~\cite{Landau:1954nau}.  Avoiding such unphysical singularities requires the introduction of an infrared (IR) confinement scale which effectively suppresses the long wavelength modes~\cite{Gribov:1977wm, Cornwall:1981zr}, or some IR regulator by adding new terms to the coupling~\cite{Shirkov:1996cd, Deur:2016tte}:  Analytic couplings  often lead to an IR fixed-point~\cite{Ayala:2014qea}.  A unified and consistent treatment of $\alpha_s$ in both the perturbative and nonperturbative domains  should incorporate confinement dynamics,  since it implies the transition from the short distance partonic degrees of freedom to the large distance hadronic degrees of freedom, the asymptotic states, where perturbative QCD is not applicable. In practice an effective description is adopted by the choice of an effective coupling subject to specific constraints.

In quantum field theory (QFT) there is some freedom in the definition of the running coupling~\cite{Deur:2016tte, Deur:2023dzc}. Several definitions restore the coupling to the status of an observable quantity, notably the  effective charge introduced by Grunberg~\cite{Grunberg:1980ja, Grunberg:1982fw}. It defines a QCD effective coupling $\alpha_{\rm eff}$ by the approximant of any given observable truncated to first order in the perturbative coupling $\alpha_s$. This makes the effective charge definition of the QCD coupling analogous to that of QED~\cite{Gell-Mann:1954yli}. An effective charge defined from a specific observable can be used for any process using the relations between the different effective charges~\cite{Brodsky:1994eh}, which guarantees the predictability of the theory.  Crucially, since the leading order approximant of an observable is scheme- and gauge-independent, its associated coupling is an observable as well, and therefore amenable to analytical continuation. An important advantage of $\alpha_{\rm eff}$ is that confinement forces and parton correlations are folded into its definition, thus both the IR and ultraviolet (UV) domains are incorporated.

In Ref.~\cite{Brodsky:2010ur}, $\alpha_{\rm eff}$ was derived in the IR domain using HLFQCD, and in Refs~\cite{Deur:2014qfa, Deur:2016cxb, Deur:2016opc} we introduced a matching procedure between HLFQCD and perturbative QCD. The procedure matches the two expressions for $\alpha_s$ at a single point $Q_0$ and yields an accurate prediction of $\alpha_s$ in agreement with both the IR data~\cite{Deur:2005cf, Deur:2008rf, Deur:2022msf} and the combined world average at the $Z^0$ mass~\cite{ParticleDataGroup:2022pth}. Yet, as we will discuss below, this prescription has its limitations.  In contrast, we use analytic continuation in this article to extend the applicability of the nonperturbative holographic QCD description of the strong coupling into the transition, near-perturbative, regime where perturbative effects become important.  Both the IR and the UV domains are incorporated into a single analytic expression providing a continuous transition between both domains for the $\beta$-function and any higher derivative.  In the present approach, the IR scale is required for the suppression of the unphysical Landau pole. In fact, the flow of singularities in the complex plane is responsible for the splitting of the Landau pole into two complex conjugate singularities.  Furthermore,  by excluding the unphysical region in coupling space, we obtain a specific relation between the mass scale underlying color confinement and the QCD scale underlying the perturbative interactions of quarks and gluons. The relation between the IR and UV scales becomes unique and leads to a single mass scale by requiring maximal analyticity in the flow of singularities.  The resulting model is compared with the experimental results for the effective charge in the  $g_1$ scheme, $\alpha_{g_1}$, defined by the Bjorken sum rule~\cite{Bjorken:1966jh, Bjorken:1969mm},  which is well-measured in both the IR and UV domains.

Our approach has bearings on the analytic S-matrix bootstrap and duality concepts and tools~\cite{Chew:1961ev, Eden:1966dnq, Veneziano:1968yb}  which arise from general principles of QFT such as symmetry, causality, unitarity and crossing, thus,  rather independently of the specific dynamics. Together with some large distance input to limit the possible solutions, this approach can lead to bounds on coupling constants~\cite{Mizera:2023tfe}.   Interestingly, in the pre-QCD period analytic continuation provided the connection between perturbation theory and strong interaction physics~\cite{Mandelstam:1959bc}.

%%%%%%%%%%%%%%

\vspace{10pt}

{\it Interpolating effective coupling.}--  In nonperturbative models, a scale is introduced to compute the hadron spectrum, whereas in a conformal gauge theory a scale, denoted by $\Lambda$ for QCD, is introduced because of the necessity of perturbative renormalization. A challenge is thus  to understand how these two scales are related depending on the symmetries of the underlying full theory. As general constraints required to extend the holographic strong coupling~\cite{Brodsky:2010ur} beyond the IR, we demand  the vanishing of the $\beta$-function in the deep IR and UV to reflect the underlying conformal symmetry of QCD and the analyticity of  $\alpha_{\rm eff}$ for all possible physical solutions consistent with the  continuity of the physical observables.

The large $N_C$ limit~\cite{tHooft:1973alw} underlies the holographic approach to strongly coupled QFT. In fact, the AdS/CFT duality, the correspondence between a classical gravity theory in a five-dimensional anti-de Sitter (AdS) space and  a conformal field theory  (CFT) in physical space-time,  corresponds to the large $N_C$ limit of a conformal theory in the  UV asymptotic AdS boundary~\cite{Maldacena:1997re, Gubser:1998bc, Witten:1998qj}. This should remain true if we modify the IR region of AdS space to introduce confinement. Our approach to holographic QCD in the light-front (HLFQCD) is based on the embedding of Dirac’s relativistic front form of dynamics~\cite{Dirac:1949cp} into AdS space. It leads to relativistic boost-invariant wave equations in physical space-time, similar to the Schr\"odinger equation in atomic physics~\cite{deTeramond:2008ht, Brodsky:2014yha}. In HLFQCD, an IR mass scale can be introduced from an emerging superconformal symmetry~\cite{deAlfaro:1976vlx, Fubini:1984hf, Akulov:1983hjq}  by constructing a scale-deformed supercharge operator, which is a superposition of supercharges within the extended graded algebra~\cite{Fubini:1984hf}. This procedure determines uniquely the confining interaction~\cite{Brodsky:2013ar, deTeramond:2014asa, Dosch:2015nwa}, as well as the corresponding modification of AdS space in the IR domain~\cite{Dosch:2016zdv, deTeramond:2023qbo}, while keeping the action conformal invariant~\cite{deAlfaro:1976vlx}: It also leads to hadronic supersymmetry between mesons, baryons and tetraquarks~\cite{Dosch:2015nwa}.

To build a specific model, we demand that in the deep IR and UV domains the $\beta$-function vanishes, thus reflecting the underlying conformal symmetry of QCD. In fact, it is clear from a physical perspective, that in a confining theory, gluons and quarks have a maximal wavelength, and thus all vacuum polarization corrections to the gluon self-energy must decouple: An infrared fixed point is a natural consequence of confinement~\cite{Brodsky:2007hb}.

Our expression for $\alpha_{\rm eff}$  incorporates the holographic results in Ref.~\cite{Brodsky:2010ur}, compatible with superconformal symmetry, and it is consistent with measurements of the strong coupling in the IR~\cite{Deur:2022msf}.  In the deep UV we will recover the leading perturbative QCD logarithmic dependence. As for the analytic continuation, we build a model, which leads to the splitting of the Landau pole into two complex conjugate singularities, thus transforming the unphysical singularities in the real axis into a singularity flow in the complex $Q^2$-plane.

We express $\alpha_{\rm eff}$ in the space-like domain by
\begin{align} \lb{alphaeff}
\alpha_{\rm eff}(Q^2) = \alpha_{\rm eff}(0) \exp\left[-\bigintssss_0^{Q^2} \frac{d u}{4 \kappa^2 + u \ln\left(\frac{ u}{\Lambda^2}\right)} \right],
\end{align}
which satisfies the constraints described above. Here, $Q^2=-q^2$ is the virtuality scale, $\kappa$ is the IR mass scale in holographic QCD, $\Lambda$ is the QCD scale governing the logarithmic evolution of the coupling in the UV, and $\alpha_{\rm eff}(0)$ is an IR fixed point value. In the $g_1$ scheme used here, it is fixed to $\pi$ by simple kinematical constraints.

Eq.~\req{alphaeff} constrains both the deep IR and UV's behavior:
\begin{align} \lb{eq:limits}
\alpha_{\rm eff}(Q^2)   \to   e^{- Q^2 /  4 \kappa^2 },  \quad  \mbox{for}  \quad    Q^2 \ll 4 \kappa^2,
\end{align}
the Gaussian  expression found in~\cite{Brodsky:2010ur}, and
\begin{align} \lb{eq:limits}
\alpha_{\rm eff}(Q^2)   \to  \frac{1}{\ln\left(Q^2 / \Lambda^2 \right)},  \quad \mbox{for}  \quad  Q^2 \gg \ 4 \kappa^2,
\end{align}
the leading logarithmic $Q^2$-dependence. Thus, Eq.~\req{alphaeff} provides a continuous transition between the IR and the UV perturbative domains.

%%%%%%%%%%%%%%%
\begin{figure}[h] 
\includegraphics[width=8cm]{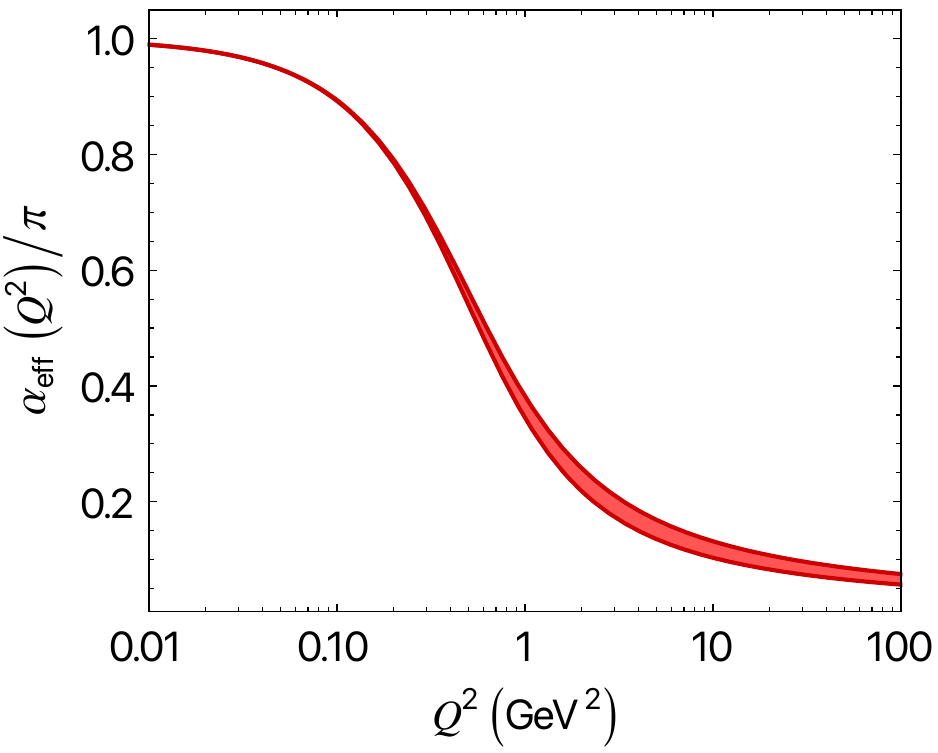}
\includegraphics[width=8.2cm]{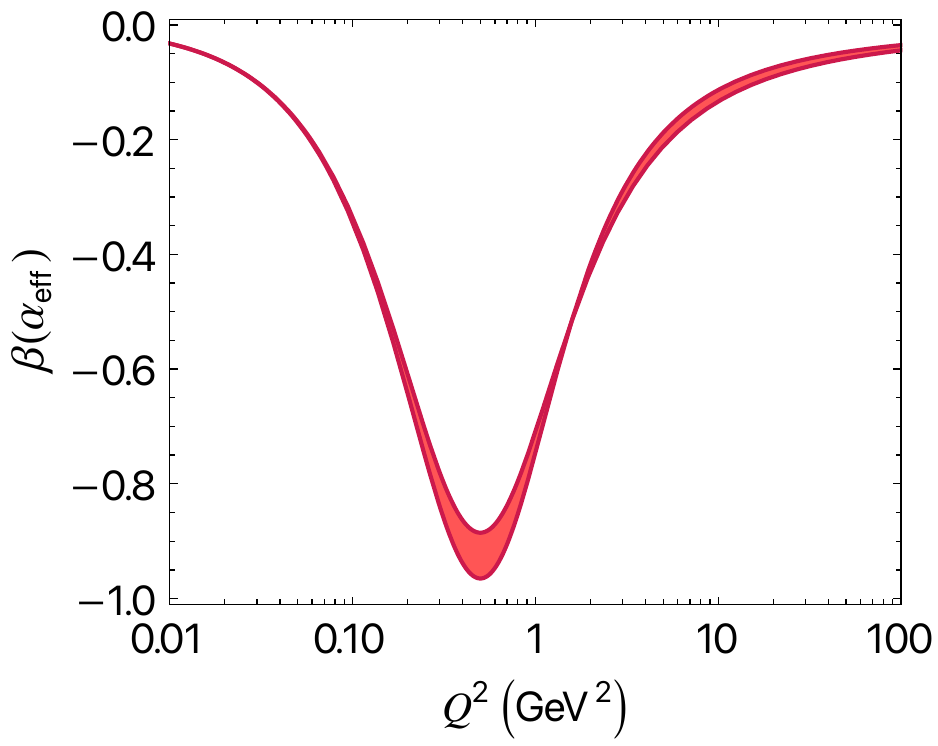}
\caption{\lb{alphabetafig} Behavior of the effective coupling $\alpha_{\rm eff}$ from Eq.~\req{alphaeff} and its $\beta$-function~\req{beta} for the range of values of $\Lambda^2 = \frac{8}{\pi} \kappa^2$ to  $\Lambda^2 = 2 \kappa^2$. The turning point of the $\beta$-function is located at $Q^2 = 2 \kappa^2$. The curves correspond to $\kappa \simeq$~0.5 GeV~\cite{Brodsky:2014yha}, determined from the hadronic mass spectrum, and the constraint $\alpha_{\rm eff}(0) = \pi$  in the $g_1$ scheme.}
\end{figure}
%%%%%%%%%%%%%%%

The $\beta$-function
\begin{align} \lb{beta}
\beta(\alpha_{\rm eff}) &=  Q^2 \,  \frac{d \alpha_{\rm eff}(Q^2)}{d Q^2} \nn   \\
              &=  -  \frac{Q^2}{4 \kappa^2 + Q^2 \ln\left(\frac{Q^2}{\Lambda^2}\right)} \, \alpha_{\rm eff}(Q^2),
\end{align} 
has a minimum at the  turning point  $Q^2 =  2 \kappa^2$, independent of the value of $\Lambda$, and vanishes in the deep IR and UV limits.  This property is illustrated in Fig.~\ref{alphabetafig}, where we choose as examples the values  $\Lambda^2 = 2 \kappa^2$ (the inflection point) and  $\Lambda^2 = \frac{8}{\pi} \kappa^2$ (the value for maximal analyticity found below).   The analytic expression~\req{beta} evolves continuously from the IR to the UV functional forms of the $\beta$-function, namely from the nonperturbative expression  in holographic QCD~\cite{Brodsky:2010ur} to the leading perturbative expression~\cite{Deur:2023dzc}
\begin{align}
\beta(\alpha_{\rm eff})  & \to  -  \frac{Q^2}{4 \kappa^2 } \, \alpha_{\rm eff}(Q^2),    \quad   Q^2 \ll 4 \kappa^2,  \\
\beta(\alpha_{\rm eff})  &  \to  - \alpha^2_{\rm eff}(Q^2),   \quad Q^2 \gg \ 4 \kappa^2.
\end{align} 
The behavior of the effective coupling $\alpha_{\rm eff}$ and its all $\beta$-function are shown in Fig.~\ref{alphabetafig} for different ratios of the scales $\kappa$ and $\Lambda$.

The large $N_C$ limit should remain valid when we extend $\alpha_{\rm eff}$ to the transition domain. This constraint implies no flavor dependence in Eq.~(\ref{alphaeff}) since in this  limit  the ’t  Hooft  coupling $\lambda_N = g^2  N_C$, for fixed $\lambda_N$, becomes independent of the number of flavors~\cite{tHooft:1973alw}. The effective coupling $\alpha_{\rm eff}$ in~\req{alphaeff}  is thus valid in the nonperturbative and  the near-perturbative domains. However, it remains  an approximation in the deep, fully perturbative, UV region, since there are no quark mass thresholds in the conformal limit on which the present derivation is based.

\vspace{20pt}

 {\it Singularity flows and confinement}-- General considerations of symmetry and analyticity have guided us to the expression of $\alpha_{\rm eff}$  given by Eq.~\req{alphaeff}. Analytic continuation of this expression leads us to study the flow of singularities in the complex $Q^2$-plane which imposes strict constraints on possible solutions.  The actual flow follows from the equation 
\begin{align} \lb{QkaLa}
4 \kappa^2 + Q^2 \ln \left(\frac{Q^2}{\Lambda^2} \right) = 0,
\end{align}
which determines the singularities of $\alpha_{\rm eff}(Q^2)$ \req{alphaeff} and its $\beta$-function \req{beta} in the space-like domain $Q^2 \ge 0$. Its solutions are given by
\begin{align} \lb{Qp}
 Q_u^2({\kappa^2}) = \Lambda^2 \exp\left[W_ 0 \left(- \frac{4 \kappa^2}{\Lambda^2}\right)\right], 
\end{align}
and
\begin{align} \lb{Qm}
 Q_l^2({\kappa^2}) = \Lambda^2 \exp\left[W_ {-1} \left(- \frac{4 \kappa^2}{\Lambda^2}\right)\right], 
\end{align}
for fixed $\Lambda$, where $u$ and $l$ refer, respectively, to the solutions in the upper and lower $Q^2$ half-planes.  $W_k(z)$, the Lambert function, is a multivalued function with a branch for each positive and negative integer value of $k$. It is also useful to express $\kappa$ as a function of $Q^2$ along  the locus of the solutions of~\req{QkaLa}
\begin{align} \lb{kaLaQ}
\kappa^2(Q^2) = - \frac{1}{4} Q^2 \ln \left(\frac{Q^2}{\Lambda^2} \right),
\end{align}
with $Q^2 \to Q_{u, l}^2$.

In the limit $\kappa \to 0$ we find the solutions
\begin{align} \lb{eq:Qp}
 Q_u^2(\kappa^2 = 0) = \Lambda^2, \quad {\rm and} \quad  Q_l^2(\kappa^2 = 0) = 0.
\end{align}
The first solution, $Q_u$, which corresponds to the principal branch of the Lambert function with $W_0(0) = 0$, gives the Landau singularity at $Q^2 = \Lambda^2$, and the second solution,   $Q_l$, which corresponds to a lower branch,  leads to an additional singularity at $Q^2 = 0$ since  $W_{-1}(0) = - \infty$.

 \begin{figure}[h] 
\includegraphics[width=8.2cm]{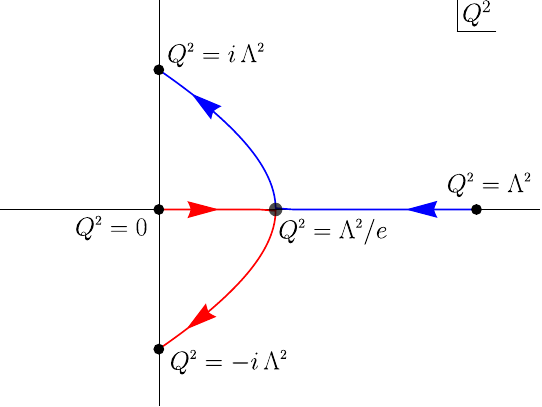}
\caption{\lb{Q2flow} 
Flow of singularities in the complex $Q^2$-plane  from the solutions of~\req{QkaLa} upon variation of $\kappa^2$ between $0 \le \kappa^2 \le \frac{\pi}{8} \Lambda^2$ for fixed $\Lambda$:  The flow paths of $Q_u$ and $Q_l$ (Eqs. \req{Qp} and \req{Qm}) are marked blue and red, respectively. The critical value  $\kappa^2 = \Lambda^2 / 4 e$ corresponds to the bifurcation point at  $Q^2 = \Lambda^2 / e$. The effective coupling~\req{alphaeff}  is not defined for $\kappa^2 < \Lambda^2 / 4 e$.  For values of $\kappa^2 > \Lambda^2 / 4 e$,  the double pole at the critical bifurcation point is split into two complex conjugate singularities, and the coupling~\req{alphaeff} can be computed for any value of $Q^2$.  The upper limit for $\kappa$ leads to the relation $\Lambda^2 = \frac{8}{\pi} \kappa^2$ between the confinement and QCD scales $\kappa$ and $\Lambda$. See also Table~\ref{Q2flowval}.}
\end{figure}

For small $\kappa > 0$ the distance between the two singularities, located initially for $\kappa = 0$, at $Q^2 = \Lambda^2$ and $Q^2 = 0$ starts to shrink until they meet at a point determined by the maximal value of $\kappa$ in the interval   $ 0 < Q^2 < \Lambda^2$ on the real axis, thus by the condition $ d\kappa^2(Q^2)/d Q^2 = 0$ with the solution $Q^2 = \frac{\Lambda^2}{e}$ as shown in Fig.~\ref{Q2flow}. Also from \req{kaLaQ} we find the critical value  $\kappa^2 = \frac{\Lambda^2}{4 e}$ (see  Table~\ref{Q2flowval}), thus the two singularities merge at the critical point
\begin{align} \lb{eq:Qp2}
Q_u^2\left(\kappa^2 = \frac{1}{4 e} \Lambda^2 \right) = \frac{1}{e} \Lambda^2, \quad {\rm and} \quad  Q_l^2\left(\kappa^2 = \frac{1}{4e} \Lambda^2 \right) = \frac{1}{e} \Lambda^2,
\end{align} 
since $W_0\left(-\frac{1}{e} \right) =  W_{-1}\left(-\frac{1}{e} \right) = -1$.  For  $\kappa^2 \le \frac{\Lambda^2}{4 e}$ the integral in Eq.~\req{alphaeff} is not defined since the singularities are located in the real axis where the integration is performed.

\begin{center} 
\begin{table}[h]
\begin{tabular}{ |c|c||c|c|} 
 \hline \hline
 $Q_u^2$ & $\kappa^2 $ &  $Q_l^2$  & $\kappa^2$ \\
 \hline 
 $\Lambda^2$ & 0 &  0  & 0\\ 
 $~\Lambda^2/e~$ & $~\Lambda^2/4e~$ & $~\Lambda^2/e~$ & $~\Lambda^2/4e~$\\ 
 $i \Lambda^2$ & $\pi \Lambda^2/8$  &  $- i \Lambda^2$ & $ \pi \Lambda^2/8$ \\ 
 \hline \hline
 \end{tabular}
 \caption{\lb{Q2flowval} Values for the solution of the flow equation~\req{QkaLa}, Eqs.~\req{Qp} and~\req{Qm}, for the specific points marked with black circles in Fig.~\ref{Q2flow}, and the corresponding values of $\kappa^2$ from \req{kaLaQ}.}
 \end{table}
\end{center}

One can combine the linear Cauchy-Riemann differential equations into the second order equations
\begin{align}
\nabla^2 {\rm Re} \, \kappa^2 = 0,  \quad  \nabla^2  \rm{Im} \, \kappa^2 = 0,
\end{align}
which implies that a maximum (minimum) of $\rm{Re}\, \kappa^2$ ($\rm{Im}\, \kappa^2$) along the real $Q^2$-axis corresponds to a  minimum (maximum) of $\rm{Re}\, \kappa^2$ ($\rm{Im}\, \kappa^2$) along the imaginary direction, thus to a saddle point, as shown in Fig.~\ref{kappa3d} (left) for ${\rm Re} \, \kappa^2$ at $\kappa^2 = \Lambda^2/4e$. We also verify in Fig.~\ref{kappa3d} (right), by taking the intersection of ${\rm Im} \, \kappa^2$ with the plane  ${\rm Im} \, \kappa^2= 0$, that, as expected, $\kappa^2$ is real on the singularity flow. 

The gradient vectors  $\vec \nabla {\rm Re} \, \kappa^2$ and $\vec \nabla {\rm Im} \, \kappa^2$  are mutually orthogonal,  $\vec \nabla {\rm Re} \, \kappa^2 \cdot \vec \nabla {\rm Im} \, \kappa^2$ = 0,  and  normal, respectively, to the two curves ${\rm Re} \, \kappa^2$ = constant and ${\rm Im} \, \kappa^2$ = constant. For both $\rm{Re} \,\kappa^2$ and $\rm{Im} \,\kappa^2$, there is an intersection between two level curves at the saddle point $Q^2 = \Lambda^2/e$, as depicted in the bottom projection planes in Fig.~\ref{kappa3d}. The singularity flow for real $\kappa^2$ ({\it i.e.,} at $\rm{Im} \, \kappa^2 = 0$) follows the vector $\vec \nabla {\rm Re} \, \kappa^2$ (Fig.~\ref{kappa3d} (left)) which, because of the orthogonality of the gradient vectors $\vec \nabla {\rm Re} \, \kappa^2$ and $\vec \nabla {\rm Im} \, \kappa^2$, should follow the level curve for $\rm{Im} \, \kappa^2$. This is verified in the projection plane in Fig.~\ref{kappa3d} (right) for the level curve which intersects the saddle point. Hence, the singularity at the critical point at $Q^2 = \Lambda^2/e$ is a bifurcation point of the singularity flow from the real axis into two complex conjugate singularities which flow into the lower and upper half-planes of the complex plane. The bifurcation point thus represents the critical value of the IR scale $\kappa^2$ above which  the coupling~\req{alphaeff} and its $\beta$-function \req{beta} -- or any higher derivative, can be computed  for any value of $Q^2 \ge 0$.

%%%%%%%%%%%%%
\begin{figure}[h] 
  \includegraphics[width=7.8cm]{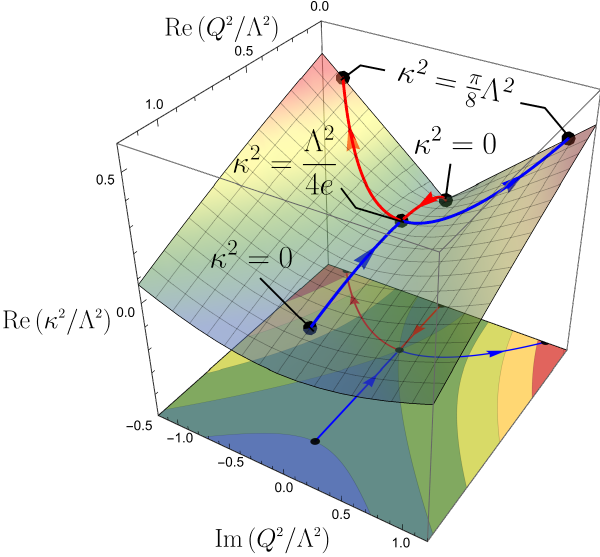}
   \hspace{0.05cm}
   \includegraphics[width=7.8cm]{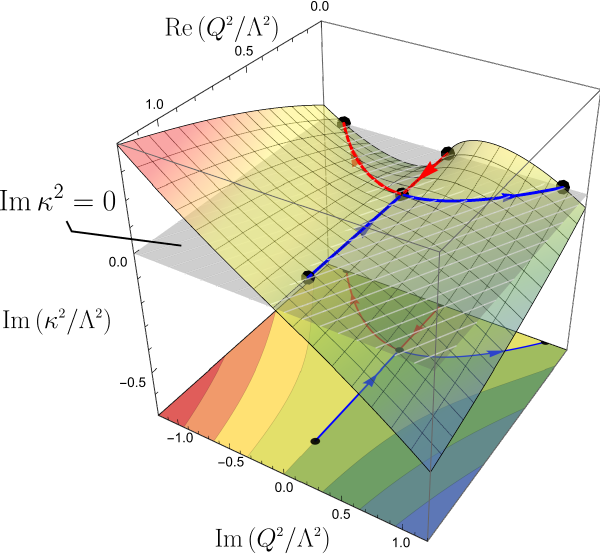}
    \caption{\lb{kappa3d} Singularity flow for the real and imaginary components of $\kappa^2$ from Eq.~\req{QkaLa}. The bifurcation point of the flow in the $Q^2$-plane is a saddle point for ${\rm Re} \, \kappa^2$ (left). The intersection of {\rm the surface} ${\rm Im} \, \kappa^2$ with the plane  ${\rm Im} \, \kappa^2= 0$,  shown in shaded grey, illustrates that  $\kappa^2$ is real along the flow (right).} 
    \end{figure}
%%%%%%%%%%%%%

By further increasing $\kappa$, the complex conjugate singularities reach their maximal separation at the points
 \begin{align} \lb{eq:Qp3}
Q_u^2\left(\kappa^2 = \frac{\pi}{8}\Lambda^2\right) =  i \Lambda^2 , \quad {\rm and} \quad  Q_l^2\left(\kappa^2 = \frac{\pi}{8} \Lambda^2 \right) = - i \Lambda^2 ,
\end{align}
which are located along the imaginary axis (see Fig.~\ref{Q2flow}) since
$W_0\left(-\frac{\pi}{2} \right) = i \frac{\pi}{2}$ and $W_1\left(-\frac{\pi}{2} \right) = - i \frac{\pi}{2}$.  From~\req{kaLaQ} we find
\begin{align} \lb{Laka}
\kappa^2 = \frac{\pi}{8} \Lambda^2,
\end{align}
for the maximal separation along the imaginary axis.   Although related by analytic continuation, the space-like and time-like domains describe physically different processes on different domains: The singularity flow from the positive-real half-plane into the negative-real half-plane in Fig.~\ref{Q2flow} is not possible. Eq.~\req{Laka}  represents an upper bound for the value of the confinement scale $\kappa$ for a given value of $\Lambda$. 

Maximal analyticity of the effective coupling in the spacelike domain leads to a unique relation between the confinement scale $\kappa$ and the QCD scale $\Lambda$: It implies that the effective coupling~\req{alphaeff} is a holomorphic function in the full $Q^2 > 0$ complex plane, except at its border $Q^2 = 0$.  This is compatible with general principles of QFT which require that spacelike observables are analytic functions in the $Q^2$-Euclidean complex plane.

\vspace{20pt} 

{\it Comparison with experiment.}-- Imposing maximal analyticity also means that the effective coupling has no free parameters once the scale $\kappa$ is fixed by an observable quantity, for example by the $\rho$ mass. Furthermore, since $\kappa$ is a physical quantity, it follows that $\Lambda$, in the present nonperturbative approach, is also an observable quantity:  The effective coupling \req{alphaeff} is thus promoted from an effective interpolation formula to an effective charge, namely to the status of an observable quantity~\cite{Grunberg:1980ja, Grunberg:1982fw}. As mentioned above in the limit of massless quarks there is no flavor dependence of the coupling. As an approximation to this limit we take the effective number of flavors to $n_f=3$. From the value $\Lambda^{(n_f=3)}_{g1} = 0.92 \pm 0.05$~GeV in the $g_1$ scheme~\cite{Deur:2016cxb}, one obtains the value $\kappa = 0.58 \pm 0.03$ GeV in the maximal analyticity limit. It is compatible with $\kappa = 0.534 \pm 0.025$ GeV from the light vector meson spectrum in HLFQCD~\cite{Sufian:2018cpj} and  $\kappa = 0.523 \pm 0.024$ GeV from the light meson and baryon spectroscopy including all radial and orbital excitations~\cite{Brodsky:2016yod}.

%%%%%%%%%%%%%%
\begin{figure}[h]
\centering
\includegraphics[width=12.8cm]{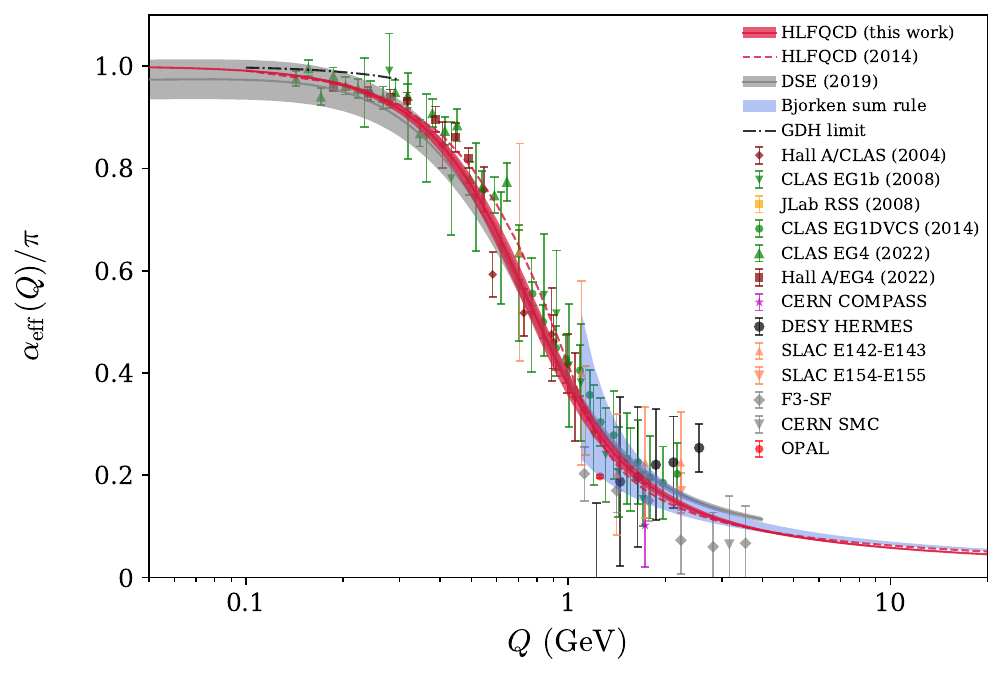}
\caption{\lb{Fig:alphaSL} 
Comparison of the effective strong coupling prediction from Eq.~\req{alphaeff} with  $\alpha_{g1}$ extracted from
the available data in Refs.~\cite{SpinMuon:1993gcv, SpinMuonSMC:1994met, SpinMuonSMC:1997voo, SpinMuon:1995svc, SpinMuonSMC:1997mkb, COMPASS:2010wkz, HERMES:1998pau, HERMES:2000apm, HERMES:2002gmr, Deur:2004ti, Deur:2008ej, Deur:2014vea, Deur:2021klh, E143:1994vcg, E143:1995rkd, E142:1996thl, E143:1995clm, E143:1996vck, E154:1997xfa, E154:1997ysl, E143:1998hbs, E155:1999pwm, E155:2000qdr, Kim:1998kia, Brodsky:2002nb}. The  red band corresponds to $\kappa = 0.534 \pm 0.025$ GeV and maximal quenching~\req{Laka}.   Here, $\alpha_{\rm eff}$ is computed in the $g_1$ scheme, which imposes $\alpha_{\rm eff}(0) = \pi$.  The dashed red curve corresponds to the point-matching procedure from Ref.~\cite{Brodsky:2010ur} for $\kappa = 0.534 ~ {\rm GeV}$.   Similar behavior has been obtained from Dyson-Schwinger and lattice computations in Refs.~\cite{Binosi:2016nme, Cui:2019dwv}, gray band. The blue band  corresponds to the perturbative 4-loop QCD prediction in the $g_1$ scheme~\cite{Deur:2022msf}.}
\end{figure} 
%%%%%%%%%%%%%%

We compare in Fig.~\ref{Fig:alphaSL} the predictions from Eq.~\req{alphaeff}, with available data, including the most recent measurements from Ref.~\cite{Deur:2022msf}. The data from the Bjorken sum, listed by experiments, are from CERN~\cite{SpinMuon:1993gcv, SpinMuonSMC:1994met, SpinMuonSMC:1997voo, SpinMuon:1995svc, SpinMuonSMC:1997mkb, COMPASS:2010wkz}, DESY~\cite{HERMES:1998pau, HERMES:2000apm, HERMES:2002gmr}, JLab~\cite{Deur:2004ti, Deur:2008ej, Deur:2014vea, Deur:2021klh} and SLAC~\cite{E143:1994vcg, E143:1995rkd, E142:1996thl, E143:1995clm, E143:1996vck, E154:1997xfa, E154:1997ysl, E143:1998hbs, E155:1999pwm, E155:2000qdr}. The data without using the Bjorken sum are from CCFR (F3-SF)~\cite{Kim:1998kia} and OPAL~\cite{Brodsky:2002nb}.  We use the value of  $\kappa= 0.534 \pm 0.025$ GeV from~\cite{Sufian:2018cpj} and the maximal quenching relation~\req{Laka}. The model describes accurately the experimental results in the nonperturbative and transition domains, consistent with a single scale governing the confinement dynamics of QCD.

\vspace{20pt}

{\it Conclusion and outlook.}--The results presented here provide a new framework for extending the gauge/gravity results for the effective coupling~\cite{Brodsky:2010ur} into the transition regime between a strongly coupled regime and an asymptotically free gauge theory.  Instead of matching the IR and UV expressions at a given point~\cite{Deur:2014qfa, Deur:2016cxb, Deur:2016opc}, we introduce a single analytic expression which leads to a continuous transition from the IR to the UV expressions of the effective strong coupling $\alpha_{\rm eff}$ and its $\beta$-function. The procedure, based on analytic continuation, is compatible with  the underlying conformal symmetry properties of QCD and its large $N_C$ constraints. The study of the analytic properties of the model determines the flow of singularities in the complex plane leading to well-defined relations between the nonperturbative and perturbative scales for the possible solutions of the strong coupling.  The Landau singularity in the real axis  splits into two complex singularities giving rise to the physical solutions. Imposing maximal analyticity in the complex $Q^2$-plane  provides a unique relation between the IR and UV scales of the strong force, leading to an accurate description of the available data in the nonperturbative and the perturbative transition region.

Extension of the present results to the full perturbative domain requires one to account for the flavor thresholds at larger scales. In the framework of the model developed here, this may be achieved by allowing for the $Q^2$ dependence of the confinement scale $\kappa$. Such a scale dependence is also required to describe the spectroscopy of heavy quark systems in holographic QCD~\cite{Gutsche:2012ez, Dosch:2016zdv, Nielsen:2018ytt}. The framework introduced in this article also naturally lends itself to the analytic continuation of our results to the entire complex plane including the timelike domain $Q^2<0$.  These subjects are under consideration.

\vspace{20pt}

{\it Acknowledgments.}-- This work is supported in part by the Department of Energy Contract No. DE-AC02-76SF00515.
It is also supported in part by the U.S. Department of Energy, Office of Science, Office of Nuclear Physics under Contract No. DE-AC05-06OR23177. T.L. is supported by the National Natural Science Foundation of China under Grants No. 12175117 and No. 12321005 and Shandong Province Natural Science Foundation Grant No. ZFJH202303. R.S.S. is supported by the Special Postdoctoral Researchers Program of RIKEN and RIKEN-BNL Research Center.

%%%%%%%%%%%%%%%%%%
\bibliography{alpha_eff_interpol}
%%%%%%%%%%%%%%%%%%

\end{document}